\documentclass[%
 aip,
 jmp,%
 amsmath,amssymb,
 reprint,%
]{revtex4-1}

\usepackage{graphicx}
\usepackage{dcolumn}
\usepackage{bm}
\usepackage[export]{adjustbox}
\usepackage{lipsum}

\newcommand*{\citen}[1]{%
 \begingroup
 \romannumeral-`\x 
 \setcitestyle{numbers}%
 \cite{#1}%
 \endgroup   
}

\begin{document}

\preprint{AIP/123-QED}

\title[Manuscript in preparation for Applied Physics Letters]{CMOS compatible W/CoFeB/MgO spin Hall nano-oscillators with wide frequency tunability}

\author{M. Zahedinejad}
\affiliation {Department of Physics, University of Gothenburg, 412 96 Gothenburg, Sweden.}
\author{H. Mazraati}
\affiliation{NanOsc AB, Kista 164 40, Sweden}
\affiliation{Department of Applied Physics, School of Engineering Sciences, KTH Royal Institute of Technology, Electrum 229, SE-16440 Kista, Sweden}
\author{H. Fulara}
\affiliation {Department of Physics, University of Gothenburg, 412 96 Gothenburg, Sweden.}
\author{J. Yue}
\affiliation {Department of Physics, University of Gothenburg, 412 96 Gothenburg, Sweden.}
\author{S. Jiang}
\affiliation{Department of Applied Physics, School of Engineering Sciences, KTH Royal Institute of Technology, Electrum 229, SE-16440 Kista, Sweden}
\author{A. A. Awad}
\affiliation {Department of Physics, University of Gothenburg, 412 96 Gothenburg, Sweden.}
\author{J. \AA kerman}
\email{johan.akerman@physics.gu.se.}
\affiliation {Department of Physics, University of Gothenburg, 412 96 Gothenburg, Sweden.}
\affiliation{NanOsc AB, Kista 164 40, Sweden}
\affiliation{Department of Applied Physics, School of Engineering Sciences, KTH Royal Institute of Technology, Electrum 229, SE-16440 Kista, Sweden}

\date{\today}

\begin{abstract}
We demonstrate low-operational-current W/Co$_{20}$Fe$_{60}$B$_{20}$/MgO spin Hall nano-oscillators (SHNOs) on highly resistive silicon (HiR-Si) substrates. Thanks to  
a record high spin Hall angle of the $\beta$-phase W ($\theta_{SH}$ = -0.53), a very low threshold current density of 3.3 $\times$ 10$^{7}$ A/cm$^2$ can be achieved. Together with their very wide frequency tunability (7--28~GHz), promoted by a moderate perpendicular magnetic anisotropy, 
this makes HiR-Si/W/CoFeB based SHNOs potential candidates for wide-band microwave signal generation. Their CMOS compatibility 
offers a promising route towards the integration of spintronic microwave devices with other on-chip semiconductor microwave components. 
\end{abstract}

\pacs{Valid PACS appear here}
\maketitle

%

The phenomenon of spin Hall effect\cite{hirsch1999prl,sinova2015rmp} 
has been widely exploited to generate 
a uniform pure transverse spin current density ($J_s$) from a lateral charge current density ($J_c$) using a non-magnetic 
metal (NM) with high spin-orbit interaction. 
The generated $J_s$ can diffuse into an adjacent ferromagnetic (FM) layer to apply spin orbit torque\cite{liu2012science} (SOT). 
Spin Hall nano-oscillators\cite{liu2012prl, awad2016nph, mazraati2016apl, durrenfeld201720,chen2016ieeerev} (SHNOs) rely on SOT to overcome the local spin wave damping and sustain 
steady-state, current- and field-tunable, precession of the magnetization around an effective magnetic field. 
SHNOs have been demonstrated using 
different FM/NM combinations 
such as NiFe/Pt\cite{demidov2012ntm,demidov2014apl,durrenfeld201720,awad2016nph}, NiFe/W\cite{mazraati2016apl}, Co$_{40}$Fe$_{40}$B$_{20}$/Pt\cite{ranjbar2014ieeemag} and YIG/Pt\cite{Collet2016ntc}. 
Finding materials with a high charge-to-spin current conversion, quantified by the 
spin Hall angle ($\mathit{\theta_{SH}}=J_s/J_c$), has been of particular interest, with 
Pt, Ta\cite{liu2012science}, W\cite{mazraati2016apl}, AuTa\cite{laczkowski2017arxiv}, and CuBi\cite{Niimi2012} being prominent examples. 

For practical applications, it is crucial that the 
NM/FM stack can be grown with high quality on CMOS compatible substrates. Ideally, the substrate should also have a high thermal conductivity to dissipate the relatively large local heat generated by the SHNO, and be inert to the NM layer at temperatures typically encountered during processing. For example, SHNO applications will likely require the use of CoFeB/MgO/CoFeB based magnetic tunnel junction (MTJ) stacks, which require an anneal step. 
The commonly used 
Si/SiO$_2$ substrate suffers from poor thermal conductivity of SiO$_2$ (1.3 W/m$\cdot$K). 
Therefore, other substrates were chosen for SHNOs, such as sapphire\cite{awad2016nph,durrenfeld201720,mazraati2016apl}, which has a rather high thermal conductivity coefficient (25 W/m$\cdot$K) 
and low microwave losses; however, sapphire is not a CMOS compatible substrate. 

High resistivity Si could be an alternative substrate as it offers both  high thermal conductivity (130 W/m$\cdot$K), five times higher than sapphire and 100 times that of Si/SiO2, while also exhibiting low microwave losses
\cite{reyes1996ieee}. However, most of the typical metallic materials used in SHNOs, such as Pt\cite{cohen1982jap}, Pd\cite{murarka1995inter}, Ni\cite{luo2010apl}, Cu, Au, and their alloys, form low-resistivity M$_{x}$Si$_{y}$ silicides at elevated temperatures, and hence would shunt the SHNO current, or inter-diffuse into the substrate, which is fatal for the CMOS process. 
Tungsten, on the other hand, can stand very high temperatures up to 800~$^{\circ}$C without forming a silicide\cite{siegal1989jap} and is already widely used  as a diffusion barrier in CMOS processes. 

In this study, we therefore demonstrate SHNOs based on  $\beta$-W/Co$_{20}$Fe$_{60}$B$_{20}$/MgO stacks fabricated on high-resistivity silicon (HR-Si) substrates.  Thanks to the very high spin Hall angle, $\mathit{\theta_{SH}}$=-0.53, 
these CMOS compatible SHNOs exhibit a very 
low auto-oscillation threshold current accompanied by a very wide current- and field-tunable microwave frequency. 
As STT-MRAM technology is pushed towards being embedded in fully depleted silicon on insulator (FD-SOI), which is widely used in CMOS\cite{doris2016sse} and CMOS RF technology\cite{shahidi2002ibm}, our demonstration is also a proof-of-concept of how SHNOs can be embedded in 
FD-SOI.

\begin{figure}[t]
    \centering
\includegraphics[trim=0cm 0cm 0cm 0cm, clip=true,width=3.4in]{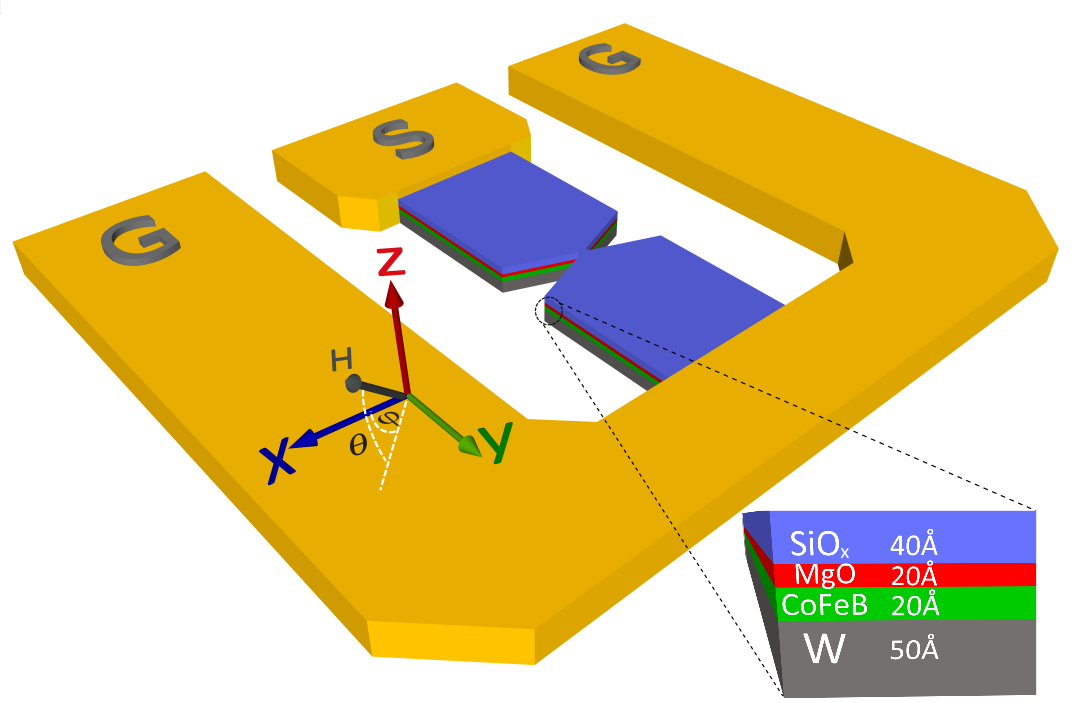}
\caption{\label{fig1} Schematic of the SHNO device including the CPW. Positive direct current flows from the signal pad to ground in the $y$-direction. $\phi$ and $\theta$ are the in-plane (IP) and out-of-plane (OOP) field angles, respectively. Inset shows the layer order and thicknesses.}
\end{figure}

The W(5)/Co$_{20}$Fe$_{60}$B$_{20}$(2)/MgO(2) stack (thicknesses in nm) 
was deposited using an AJA Orion-8 magnetron sputtering system working at a base pressure of 1$\times$10$^{-8}$ mTorr while the Argon pressure during sputtering was kept at 3~mTorr for all the layers. dc and rf sputtering were used for the depositions of metallic and insulting layers, respectively. As 
it is crucial to keep the W deposition rate low to obtain the $\beta$-phase with a high spin Hall angle\cite{mazraati2016apl,zhang2016apl}, it was kept at 0.09~\AA/s; the same rate was used for 
Co$_{20}$Fe$_{60}$B$_{20}$, 
while that of the MgO layer was 0.04~\AA/s. The stack was deposited on intrinsic HR-Si with $\rho_{Si}>$10 k$\Omega \cdot$cm.
 
To crystallize the Co$_{20}$Fe$_{60}$B$_{20}$ layer 
the stack was annealed 
at 300~$^{\circ}$C for 60 minutes at the system base pressure. 
The crystallization starts at the  
interface to MgO while W serves as a Boron getter\cite{lee2016npg} to mediate the Co$_{20}$Fe$_{60}$B$_{20}$ crystallization. The annealing 
was followed by deposition of 40~\AA~of SiO$_x$ to protect the MgO layer from reacting with the ambient moisture. The $\beta$-phase W and Co$_{20}$Fe$_{60}$B$_{20}$ 
resistivities were measured to be 200 $\mu \Omega \cdot$cm and 90 $\mu \Omega \cdot$cm, respectively. Obtaining high resistivity W layer is a sign of having either mixed $\alpha + \beta$ or $\beta$-rich phase W as it has been studied extensively in literature \cite{mazraati2016apl,zhang2016apl,mondal2017prb,Demasius2016,pai2012apl}. For the sake of notation, We use the term $\beta$-phase for both cases for the rest of the manuscript.

In order to fabricate nano-constrictions, the sample surface was covered with negative electron resist followed by electron beam lithography (JEOL9300XS). 
Nano-constrictions of different widths were defined in 4$~\mu$m $\times$ 12$~\mu$m mesas. We also made 6$~\mu$m $\times$ 18$~\mu$m bars which was used to characterize the stack using spin-torque-induced ferromagnetic resonance measurements (ST-FMR). 
The patterns were transferred to the stack by Ar ion beam etching in an Oxford Ionfab 300 Plus etcher. The negative resist was subsequently removed and optical lift-off lithography was carried out to define ground--signal--ground (GSG) coplanar waveguides (CPW) of a thick Cu(950 nm)/Au(50 nm) bilayer. 
To ensure good electrical contact between the CPW and the SHNOs, the MgO/SiO$_x$ layers were removed in the CPW defined area by substrate plasma cleaning at 40 W rf power in our AJA Orion-8 sputtering machine right before Cu/Au deposition.

Figure~\ref{fig1} shows a schematic of the SHNO device including the CPW. The inset shows the magnified view of the constituent layers and their thicknesses. A positive direct current is defined to flow from the signal pad to ground in the $y$-direction. 
$\phi$  and $\theta$ are the in-plane (IP) and out-of-plane (OOP) field angles, respectively. 


\begin{figure}[b]
    \centering
\includegraphics[trim=0cm 0cm 0cm 0cm, clip=true,width=3.4in]{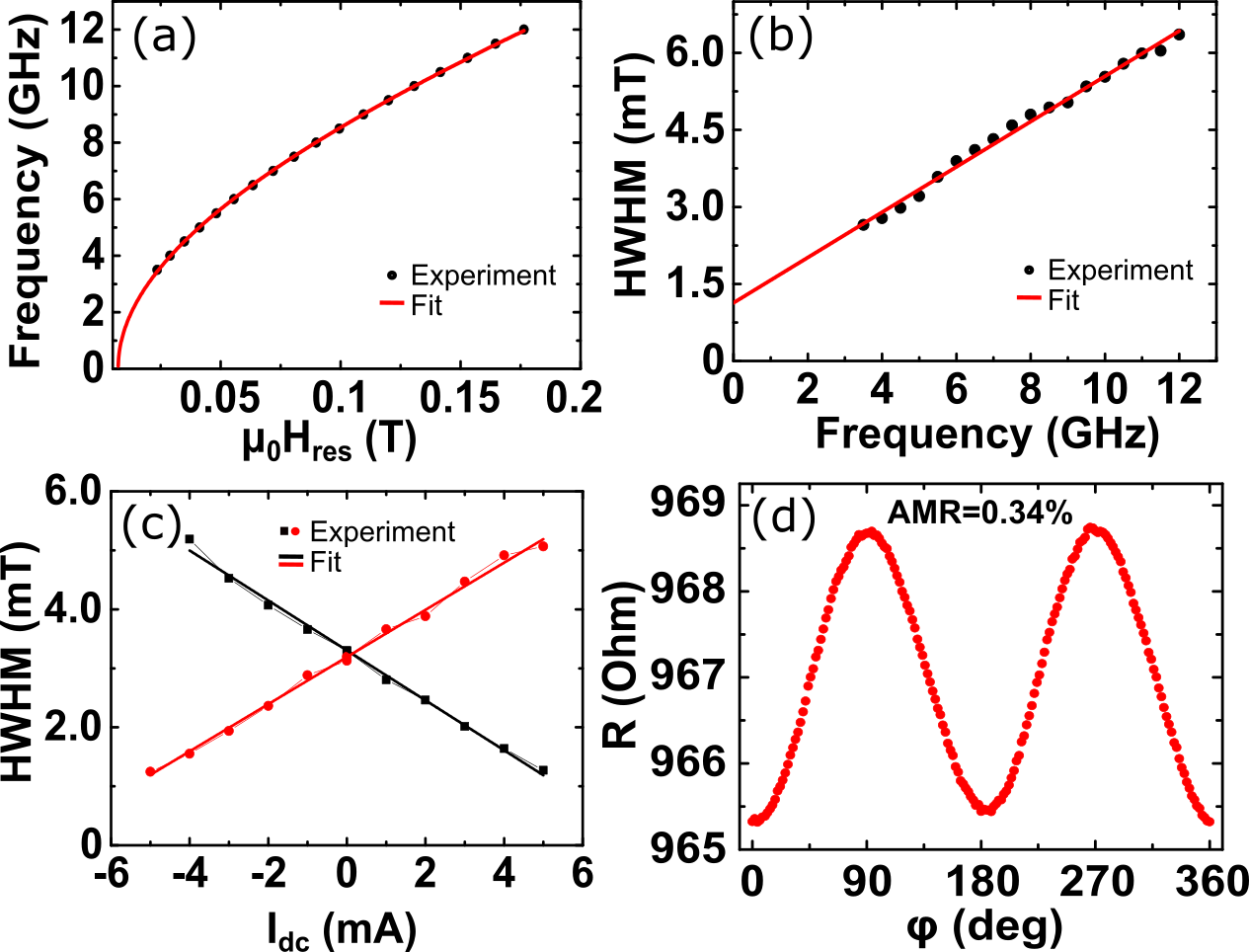}
\caption{\label{fig2}  (a) Resonance frequency vs. in-plane field from ST-FMR measurement on a 6~$\mu$m-wide bar-shaped structure. (b) ST-FMR linewidth vs.~frequency. (c) ST-FMR linewidth vs.~current in a in-plane magnetic field along $\phi$=30$^{\circ}$ (black dots) and 210$^{\circ}$ (red dots). (d) Obtained AMR for w=120 nm SHNO.}
\end{figure}

We first investigated the magnetodynamic properties of the stack using ST-FMR measurement on a 6 $\times$ 18$~\mu m^2$ bar by applying a direct current (dc) and a 98.76 Hz modulated microwave current through the input port of a bias-tee at a fix rf frequency, and detecting the resulting voltage on its output port using a lock-in amplifier while sweeping the magnetic field (0--0.2~T). The voltage response from each sweep was fit to a sum of one symmetric and one antisymmetric Lorentzian sharing the same resonance field and linewidth.\cite{liu2011prl} Figure~\ref{fig2}(a) shows the resonance fields extracted at different microwave frequencies (3--12 GHz) without any direct current. The field dependence is well described by a Kittel behavior\cite{Kittel1948} with an effective magnetization of $\mathit{M_{eff}}$ = 0.71~T and a gyromagnetic ratio of $\gamma/2\pi$ = 30.9~GHz/T. It is interesting to compare  $\mathit{M_{eff}}$ = 0.71~T with $M_S =$ 1.17 T measured using Alternating Gradient Magnetometry (AGM). According to equation $M_{eff}= M_{s}-H_{k}^{\perp}$. $H_{k}^{\perp}$, the significantly lower $\mathit{M_{eff}}$ indicates a moderate perpendicular magnetic anisotropy (PMA) field of 0.46 T.\cite{liu2011apl,peng2015sr}

Figure \ref{fig2}(b) shows the extracted linewidth for the same set of sweeps along with a linear fit~\cite{Kittel1948} yielding a low Gilbert damping of $\alpha$ = 1.36 $\times$ 10$^{-2}$. Figure~\ref{fig2}(c) shows the current-induced linewidth changes of the ST-FMR spectra due to SOT, at a fixed frequency of 5 GHz. The linewidth has a linear current dependence with a negative (positive) slope for an IP field angle of $\phi$ = 30$^{\circ}$ ($\phi$ = 210$^{\circ}$) from which we can extract $\mathit{\theta_{SH}}=$ -0.53, which is the highest reported $\mathit{\theta_{SH}}$ for W.\cite{Demasius2016,mondal2017prb,zhang2016apl} Figure~\ref{fig2}(d) shows an anisotropic magnetoresistance of 0.34$\%$ for a 120 nm wide nano-constriction. This is about an order of magnitude larger than the previously reported value of 0.026$\%$ for CoFeB \cite{ranjbar2014ieeemag} and relates to the fact that our annealing step turns the initially amorphous Co$_{20}$Fe$_{60}$B$_{20}$ into a more poly-crystalline microstructure.


\begin{figure}[t]
    \centering
\includegraphics[trim=0cm 0cm 0cm 0cm, clip=true,width=3.3in]{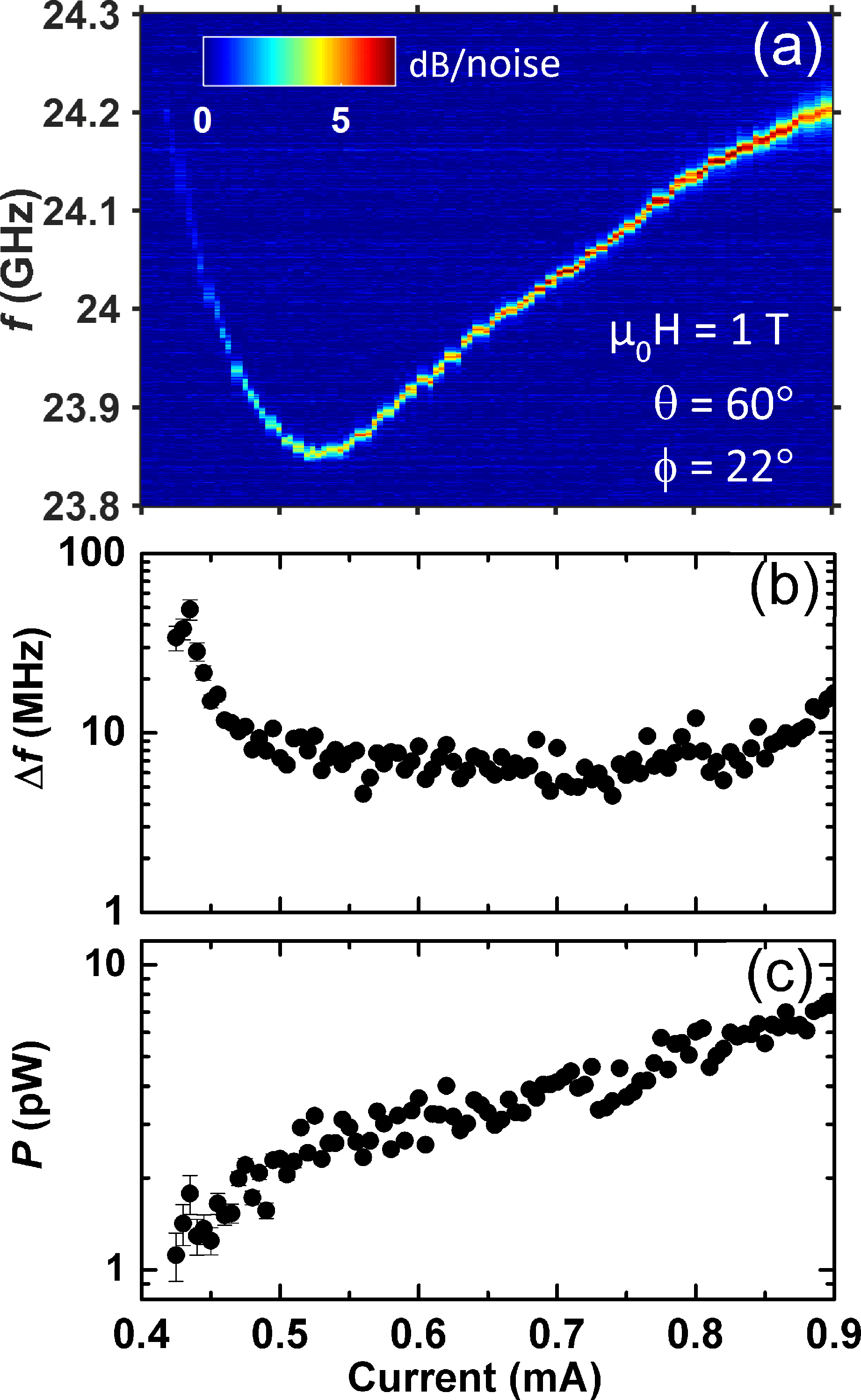}
\caption{\label{fig3} (a) Power spectral density (PSD) vs current  for a w = 120 nm SHNO in an applied magnetic field of ${\mu}_0H$=1 T along $\phi$=22$^\circ$ and $\theta$=60$^\circ$ (b) Linewidth of the extracted auto-oscillation peaks and (c) Integrated power of the extracted individual peaks.} 
\end{figure}
To investigate auto-oscillations, the lock-in amplifier was replaced with a low noise +43 dB amplifier and a spectrum analyzer, and the generated microwave power spectral density (PSD) was recorded as a function of direct current and applied magnetic field. 
Figure~\ref{fig3}(a) shows the current dependent PSD from a $w$ = 120 nm nano-constriction 
measured in a constant 1 T field with $\phi$ = 22$^\circ$ and $\theta$ = 60$^\circ$. 
The non-monotonic current dependence---first red-shifted, then blue-shifted---is typical for a nano-constriction SHNO in a close-to-perpendicular magnetic field.
\cite{dvornik2017apr,awad2016nph} The auto-oscillation region starts 
at the constriction edge (red-shift) 
and, as the current increases, moves to the center and expands to fill the constriction region (blue-shift)\cite{awad2016nph}. 
The linewidth (Figure~\ref{fig3}(b)) and integrated power (Figure~\ref{fig3}(c)) were obtained  by Lorentzian fits to the PSD; 
the linewidth decreases and the power increases with increasing current. The maximum integrated power of about 8 pW has been achieved at low operational current of 0.9 mA which is considerably higher compared to previous reports on NiFe/Pt \cite{awad2016nph,durrenfeld201720} and NiFe/W \cite{mazraati2016apl} SHNOs. 
The 
high $\mathit{\theta_{SH}}$ reduces the auto-oscillation threshold current to about $I_{th}$ = 0.5 mA (determined as in Ref. [\citen{tiberkevich2007apl}]), corresponding to a 4.4 $\times$ 10$^{7}$ A/cm$^2$ current density in the W layer, \emph{i.e.}~very close to 
previously reported values for NiFe/W SHNOs in in-plane fields\cite{mazraati2016apl}. Reducing the field magnitude to 0.8 T, we extracted an even lower threshold of about 3.3 $\times$ 10$^{7}$ A/cm$^2$.  It may be pointed out that wider nano-constriction width offers an increased auto-oscillation mode volume which in turn reduces the line-width of auto-oscillation peaks and enhances the peak power. However, lateral confinement of the nano-constriction, within the limit of reasonable yield of SHNO devices, results into a linear rise in resistance and an equivalent decrease in threshold current, thereby reduces the power consumption via quadratic dependence on current. Therefore, an optimum nano-constriction width of 120 nm has been chosen in this present work to efficiently realize the conducive operational characteristics with lesser power consumption.

\begin{figure}[b]
    \centering
\includegraphics[trim=0cm 0cm 0cm 0cm, clip=true,width=3.3in]{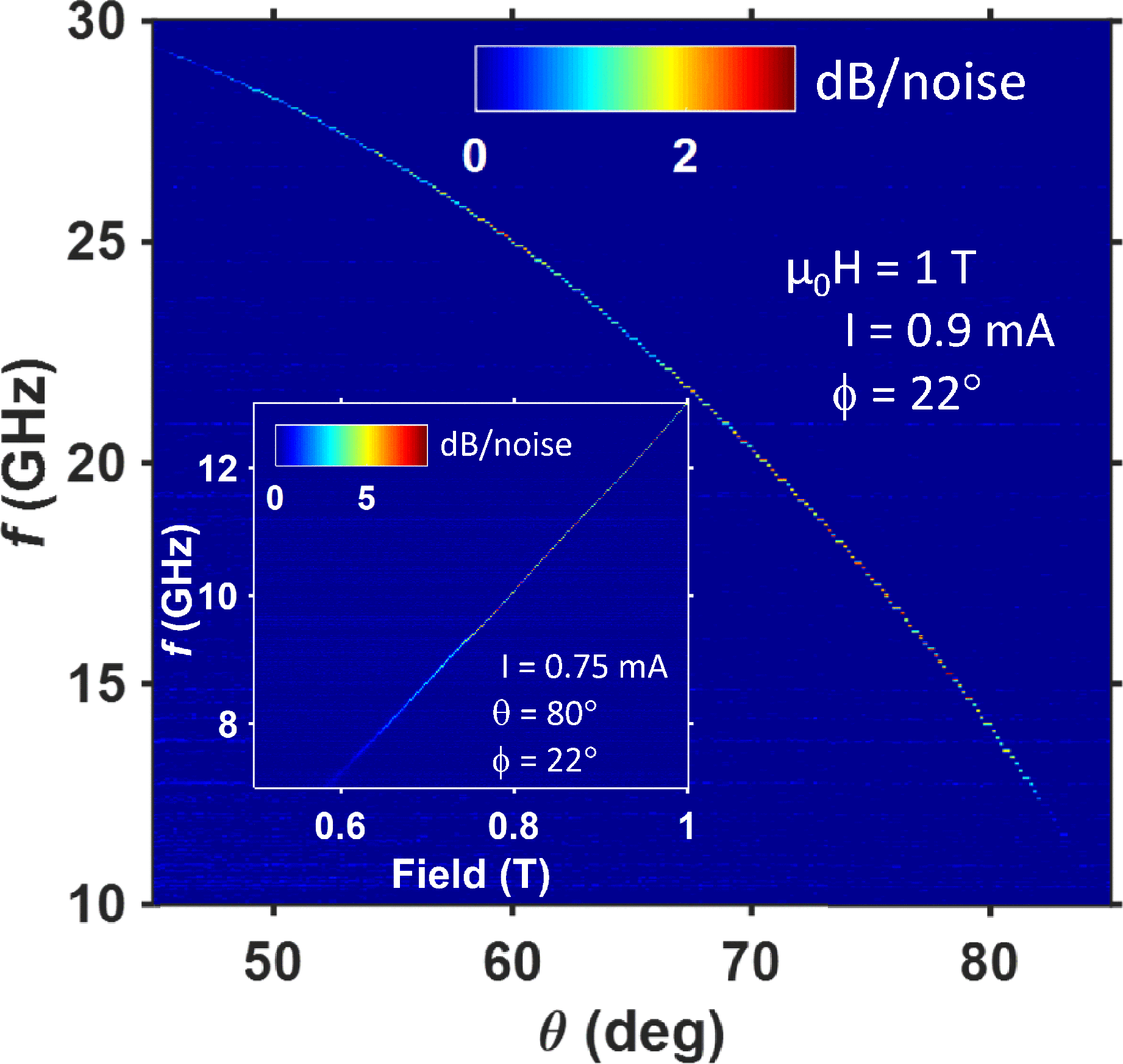}
\caption{\label{fig4} Main panel: Out-of-plane angular dependence of the microwave precession frequency at fixed $\phi$=22$^\circ$, $I$=0.9 mA and ${\mu}_0H$=1 T for a w = 120 nm SHNO. Inset displays the field dependence of microwave precession frequency for the same SHNO at fixed $I$=0.75 mA along $\phi$=22$^\circ$ and $\theta$=80$^\circ$.}
\end{figure}

To examine the operational frequency range of our SHNOs, we investigated their auto-oscillations as a function of OOP angle at a fixed field magnitude ($\mu_{0}H$ = 1 T) and current ($I_{dc}=$ 0.9 mA). Figure~\ref{fig4} displays the resulting auto-oscillations, which range from 
12 to 28 GHz. 
As the OOP angle increases, the IP component of the magnetic field becomes smaller in magnitude and therefore shifts both the FMR frequency and the auto-oscillation operating frequency to lower values. It can be noted that our SHNOs are operative in a wide frequency range from 12~GHz up to 28~GHz by just tailoring the OOP angle. The inset in Figure~\ref{fig4} further shows the field dependence of auto-oscillation frequency at a fixed OOP angle ($\theta$ = 80$^\circ$) and dc-current ($I_{dc}$=0.75 mA). The variation in the magnitude of external magnetic field gives rise to the tilting of internal magnetization angle with respect to IP direction which is manifested as a linear rise of the precessional frequency from 7~GHz up to 13~GHz. 
It is noteworthy that an observable auto-oscillation signal clearly persists down to much lower field values than that suggested by $M_S$, which  emphasizes the beneficial role of the moderate PMA in extending the field range. As a potential route to further reduce the required magnetic field to levels where integrated thin film permanent magnets would be sufficient, one could increase the PMA by thinning the CoFeB. 

In conclusion, we have demonstrated CMOS compatible nano-constriction SHNOs 
based on W/Co$_{20}$Fe$_{60}$B$_{20}$/MgO stacks grown on highly resistive Si substrates. Thanks to the record high spin Hall angle of the W layer ($\theta_{SH}$ = -0.53) and the moderate perpendicular magnetic anisotropy of the Co$_{20}$Fe$_{60}$B$_{20}$ layer, these SHNOs exhibit very low threshold currents and operate over a very wide frequency range, 
7--28~GHz. 
Also, such devices could be further integrated on Si with other microwave components to offer more compact and tunable microwave devices used in RF CMOS communication systems.

This work was supported by the Swedish Foundation for Strategic Research (SSF), the Swedish Research Council (VR), and the Knut and Alice Wallenberg foundation (KAW). This work was also supported by the European Research Council (ERC) under the European Community's Seventh Framework Programme (FP/2007-2013)/ERC Grant 307144 ``MUSTANG''.


\section{References}
%

\end{document}